\documentclass[aps,prb,floatfix,twocolumn,showpacs,showkeys,amsmath,amssymb,superscriptaddress,10pt]{revtex4-1}

\usepackage{amsthm}
\usepackage{bm}
\usepackage{epsfig}
\usepackage{verbatim}
\usepackage{graphicx}

\newcommand{\beq}{\begin{equation}}
\newcommand{\eeq}{\end{equation}}
\newcommand{\beqa}{\begin{eqnarray}}
\newcommand{\eeqa}{\end{eqnarray}}

\usepackage{xcolor}

\begin{document}

\title{Inverse transition in the dipolar frustrated Ising ferromagnet: the role of domain walls}

\author{Luciana Ara\'ujo Velasque}
\email{lu.pos.fis@gmail.com}
\affiliation{Departamento de F\'{\i}sica,
Universidade Federal do Rio Grande do Sul\\
CP 15051, 91501-970 Porto Alegre, RS, Brazil}
\author{Daniel A. Stariolo}
\email{daniel.stariolo@ufrgs.br}
\affiliation{Departamento de F\'{\i}sica,
Universidade Federal do Rio Grande do Sul and
National Institute of Science and Technology for Complex Systems\\
CP 15051, 91501-970 Porto Alegre, RS, Brazil}
\author{Orlando V. Billoni}
\email{billoni@famaf.unc.edu.ar}
\affiliation{Facultad de
Matem\'atica, Astronom\'{\i}a y F\'{\i}sica, Universidad Nacional
de C\'ordoba, Instituto de F\'{\i}sica Enrique Gaviola (IFEG-CONICET)\\ Ciudad Universitaria, 5000 C\'ordoba, Argentina}

\date{\today}

\begin{abstract}

We present a theoretical study aimed to elucidate the origin of the inverse symmetry breaking transition observed 
in ultrathin magnetic films with perpendicular anisotropy.  We study the behavior of the  dipolar frustrated
Ising model in a mean field approximation as well as two other models with simple domain walls. 
By a numerical analysis we show that the internal
degrees of freedom of the domain walls  are decisive for the presence of the inverse symmetry breaking transition. 
In particular, we show that in a sharp domain wall model the inverse transition is absent.
At high temperatures the additional degrees of freedom of the extended domain walls increase the entropy of the system 
leading to a reduction of the free energy of the stripe phase. Upon lowering the temperature the domain walls become narrow
and with the corresponding degrees of freedom effectively frozen, which eventually induces an inverse transition to the competing homogenous phase. 
We also show that, for growing external field at constant temperature, 
the stripe width grows strongly when approaching the critical field line and diverges at the transition.  
These results indicate that the inverse transition is a  continuous phase transition and that the domain
wall profiles as well as the temperature has little effect on the critical behavior of the period of the domain as
function of the applied field.
\end{abstract}


\maketitle

\section{Introduction}
\label{Intro}
Experiments on ultra-thin ferromagnetic films of Fe/Cu(001) have shown that, under a perpendicular
magnetic field, the field versus temperature phase diagram displays inverse symmetry-breaking
transitions~\cite{SaLiPo2010,SaRaViPe2010}. Without external field, these systems show low
temperature phases with modulated order in the out of plane component of the magnetization. 
These phases are a consequence of the competition between exchange, dipolar and uniaxial
anisotropies, and have been extensively studied due to the rich phenomenology~\cite{PoVaPe2003,
ChWuWoWuScDoOwQi2007,PiBiStCa2010,AbKaPoSa1995,DeMaWh2000,NiSt2007}. For weak applied fields
a finite global magnetization begins to develop, it is manifested in the appearance of an
asymmetry between the two preferential directions of the out of plane local magnetization.
At higher fields, stripe-like phases loose stability and a bubble-like phase may be present
before saturation when a uniformly magnetized phase sets in~\cite{GaDo1982,DiMu2010,CaCaBiSt2011}.
Both in experiments and in theoretical models it was found that, decreasing the temperature
at a fixed external field value, it is seen a sequence of phases from a disordered one at
high temperature to a modulated phase via symmetry breaking of translational,
rotational or both invariances in the out of plane magnetization. Interestingly, upon further lowering 
of the temperature at fixed field value, a second transition is found, which corresponds to an
inverse symmetry breaking, restoring spatial invariance at low temperatures~\cite{SaLiPo2010,SaRaViPe2010,
PoGoSaBiPeVi2010,CaCaBiSt2011}. 

Inverse symmetry breaking (ISB) has been reported many times before. The most usual cases correspond to
{\em inverse melting} or {\em inverse freezing}, see e.g. \onlinecite{ScSh2005} and references therein.
The reentrance of a more symmetric phase from an already broken symmetry one at low temperatures can
often been traced to a subtle interplay between energy and entropy, while temperature or another thermodynamic
control parameter is varied. As stated in \onlinecite{ScSh2005} {\em inverse melting happens if, and only
if, the so called ``ordered'' phase (crystal) admits more entropy than the ``disordered'' state; this may
occur, e.g., if in the liquid phase some of the degrees of freedom of the elementary constituents are
frozen, and melt in the crystalline phase}. We will see that a similar phenomenon occurs in the ultrathin
magnetic model studied in this work. In spin systems, inverse freezing has been reported mainly in theoretical 
descriptions of spin glasses and disordered models, in which frustration leads to complex entropic contributions
~\cite{CrLe2005,Se2006,PaLeCr2010,ThKa2011}. In this work, we report on a detailed analysis of the ISB
 transition in a well known model for ultrathin ferromagnetic films with perpendicular anisotropy:
the dipolar frustrated Ising model (DFIM), and unveil the nature of the inverse symmetry breaking phenomenon in
this system. In
previous work, the existence of inverse symmetry breaking was established theoretically based on a scaling 
hypothesis for modulated systems~\cite{PoGoSaBiPeVi2010} and confirmed and explored to some extent in a
coarse-grained model with a Landau-Ginzburg type effective free energy~\cite{CaCaBiSt2011}. This analysis
allowed to confirm the existence of ISB in two-dimensional models, but the phase diagram was restricted to
relatively high temperatures due to the effective nature of the model. No attempt was done to explain the
nature of the ISB. Here we show the mean field phase diagram of the DFIM in the whole temperature range where
ISB transitions can be observed. We focus our analysis on the behavior of the ``asymmetric-stripe'' solutions.
Although bubble solutions compete with stripes and may be thermodynamically
stable in some region of the $H-T$ plane, it was shown in \onlinecite{CaCaBiSt2011} that its domain of stability
is probably restricted to a small region in the high temperature sector of the $H-T$ plane. Instead, at lower
temperatures where the ISB phenomenon is observed, stripes with asymmetry in the net magnetization are the relevant solutions.
We show that asymmetric stripes display reentrant behavior. In order to elucidate the nature of the reentrance,
we analyze two other models which differ in the structure of the domain walls: a ``sharp-wall'' model, in which the
transitions between positive and negative magnetizations in the stripe patterns are abrupt, and a ``two-spin-wall''
model, in which the transitions involve two spins. These models then are compared with the full mean field model,
in which the domain walls, although much simpler than domain walls in real systems, can be spatially extended with a width that
depends on temperature and magnetic field. Our main result is that the sharp-wall model does not support reentrant
behaviour, while the two-spin-wall model does. Then we conclude that the domain wall degrees of freedom are essential to
drive the ISB transitions. We analyze in detail the energy, entropy and magnetization contributions to the free energy 
and show that the enhanced entropic contribution of domain walls in the full model are responsible for the reentrant
behaviour. Besides, we show that the three models behave in the same way at sufficiently low temperatures, as expected,
and analyze the behaviour of the stripe widths and asymmetry as a function of magnetic fields. At low temperatures our
results are compared with recent analytic work on a sharp model at $T=0$ showing a very good agreement with theoretical
predictions~\cite{JoPaGa2013}.

The organization of the paper is as follows: in Section \ref{mf} we introduce the models and discuss the mean field
solution in an external field. In \ref{assy} we analyze the solutions at very low temperatures and compare our
results with exact results known for $T=0$ in a sharp-wall model.
In \ref{phases} we present the results for the $H-T$ phase diagrams.
In Section \ref{energies} we analyze the free energy contributions at low and high temperatures. 
We conclude in Section \ref{conc} with a summary of the results and perspectives for future work.

\section{Model and Mean Field solutions}
\label{mf}
The dipolar frustrated Ising model in the two-dimensional square lattice, suitable to describe ultra-thin films 
with perpendicular anisotropy, is defined by the Hamiltonian:
\begin{equation}
{\mathcal{H}} = -\frac{1}{2} \sum_{i,j}J_{ij}S_iS_j - H\, \sum_i S_i\;,
\end{equation}
where
$\{S_i=\pm 1, i=1,\ldots ,N\}$ are $N$ Ising spins in a $L\times L=N$ two-dimensional square lattice,
\begin{equation}
 J_{ij} = \left\{ \begin{array}{ll}
 \delta-1 \; & \quad \mbox{if} \quad i,j \quad \mbox{nearest neighbors} \; \\
 0 \; & \quad \mbox{if} \quad i = j \;\\
 -\frac{1}{r_{i,j}^3}\; & \quad \mbox{otherwise} \;,
\end{array} \right.
\end{equation}
and $H$ is an homogeneous magnetic field perpendicular to the film. Because we are interested in describing thermodynamic
phases with modulated magnetization profiles one needs to consider the local magnetizations
$m_i=\langle S_i \rangle$. Then, the free energy of the model in the mean field approximation reads:
\beqa
\label{fmf}
 F_{MF} &=& -\frac{1}{2} \sum_{i,j}J_{ij}m_i m_j - H\, \sum_i m_i + \frac{k_BT}{2} \times \\
&& \sum_i \left[ (1+m_i) \ln(1+m_i) + (1-m_i) \ln(1-m_i)\right]\;.\nonumber
\eeqa
In this expression the first term on the right side  is the internal energy, the second the Zeeman term, and
the last one is the entropy.
For $H=0$ the ground state and low temperature
equilibrium states are known to correspond to stripe patterns of period $\lambda$. In this case, each period is
composed of equal $\lambda^+=\lambda^-=\lambda/2$ positive(negative) magnetization sites. At finite $H$ the direction
parallel to the field will be favored and the stripes continue to be equilibrium states but become asymmetric with
$\lambda^+ \neq \lambda^-$ and $\lambda^++\lambda^-=\lambda$. The modulation length $\lambda$ is a function of
temperature and magnetic field and then has to be considered as a variational parameter for the minimization of (\ref{fmf}).
We found more efficient to minimize directly the free energy with respect to $\lambda+1$ parameters 
using a standard numerical minimizer than solving
the equivalent set of $N$ coupled non-linear state equations $\partial F/\partial m_i=0$. Also, because of the long
range nature of the dipolar interaction, we found useful to work with the Fourier transform of the energy term in (\ref{fmf}).
The Fourier transform of the couplings in the square lattice is given by~\cite{PiCa2007}:
\beq
\label{jk}
 J_{k_x}=2\delta(\cos k_x + 1)-k_x^2 + 2\pi |k_x|-\frac{2\pi^2}{3} - 2\zeta(3)\;,        
\eeq
where we considered stripes perpendicular to the $x$ axis ($k_y=0$) and $\zeta(x)$ is the Riemann zeta function. 

In order to minimize the free energy we proceed as follows. Given a trial magnetization
profile  $\{m_j\}$ with $j=1,2, ... ,\lambda$ we compute its Fourier transform
\beq
m_{k_x} = \frac{1}{\sqrt{\lambda}}\sum_{j=1}^{\lambda}m_{j}e^{ik_xj} ,
\eeq
where the wave vectors take the values $k_x=2\pi n/\lambda$, 
with $n=-\lambda/2+1,\ldots,0,\ldots,\lambda/2$ for periodic boundary conditions.
Then using  Eq. (\ref{jk}) we compute the energy term of the free
energy per unit of length $f_{MF} = F_{MF}/\lambda$. 
\beq
u_{FM} = \frac{1}{2 \lambda} \sum_{k_x}  J_{k_x}\,|m_{k_x}|^2.
\eeq
In this way long-range dipolar interactions are taken into account.
The entropy term is computed directly using the ${m_j}$. 
Varying also the periodicity $\lambda$ we obtain the most general magnetization profile
that minimizes the free energy. 
In order to capture the influence of the domain wall structure we also analyzed two simplified
profiles: one that includes a sharp-wall and other with an extended domain wall that we call 
the two-spin-wall model~\cite{ViSaPoPePo2008}. They are defined as follows:
\subsubsection{Sharp-wall}
This is a profile with an Ising like domain wall with
\begin{equation}
 m_{j} = \left\{ \begin{array}{ll}
 m_0 \; & \quad \mbox{if} \quad j \le a  \; \\
 -m_0 \; & \quad  \mbox{otherwise.} \;
\end{array} \right.
\end{equation}
This profile has three parameters, the magnetization at the domains $m_0$, the domain wall position $a$, and the
stripe pattern period $\lambda$.
\subsubsection{Two-spin-wall}
\begin{figure}
\centering
\includegraphics[scale=0.3]{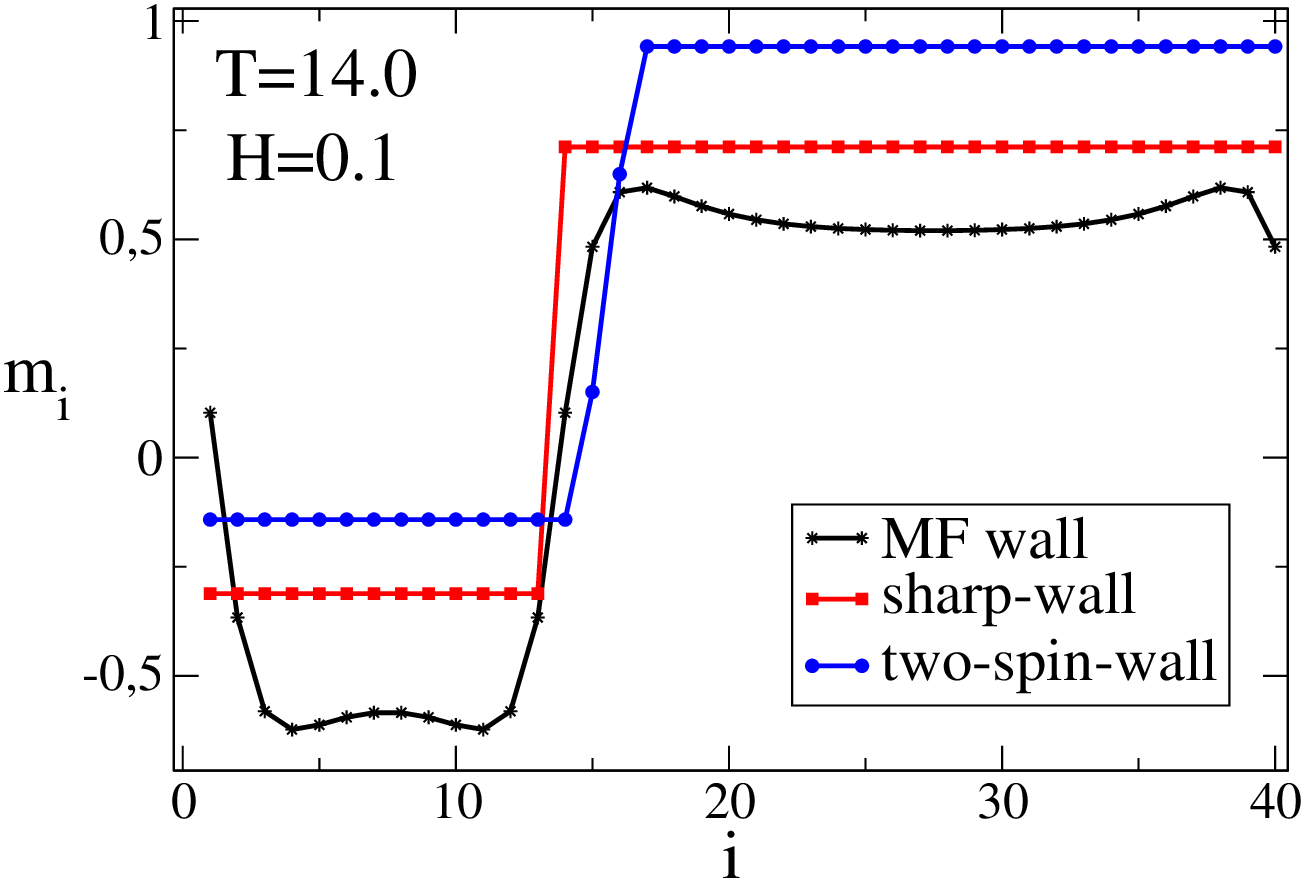}
\caption{(Color online) Magnetization profiles of the three models studied.
The amplitudes are shifted in order to illustrate better the shape of each profile.}
\label{profile}
\end{figure}

In this model the domain wall consists in a couple of spins 
which can adjust their magnetization independently of the domain magnetization:
\begin{equation}
 m_{j} = \left\{ \begin{array}{ll}
 m_0 \; & \quad \mbox{if} \quad j \le a  \; \\
 m_1 \; & \quad \mbox{if} \quad j = a+1 \; \\
 -m_1 \; & \quad \mbox{if} \quad j = a+2 \; \\
 -m_0 \; & \quad  \mbox{otherwise.} \;
\end{array} \right.
\end{equation}
This model adds a new parameter with respect to the sharp-wall model, namely the magnetization inside the domain 
wall $m_1$, i.e. this is a four parameter profile. In Figure \ref{profile} representative profiles of the three models
studied are shown. 

\section{Results}
\label{res}

\subsection{Stripe widths and asymmetry}
\label{assy}
\begin{figure}
\centering
\includegraphics[scale=0.3]{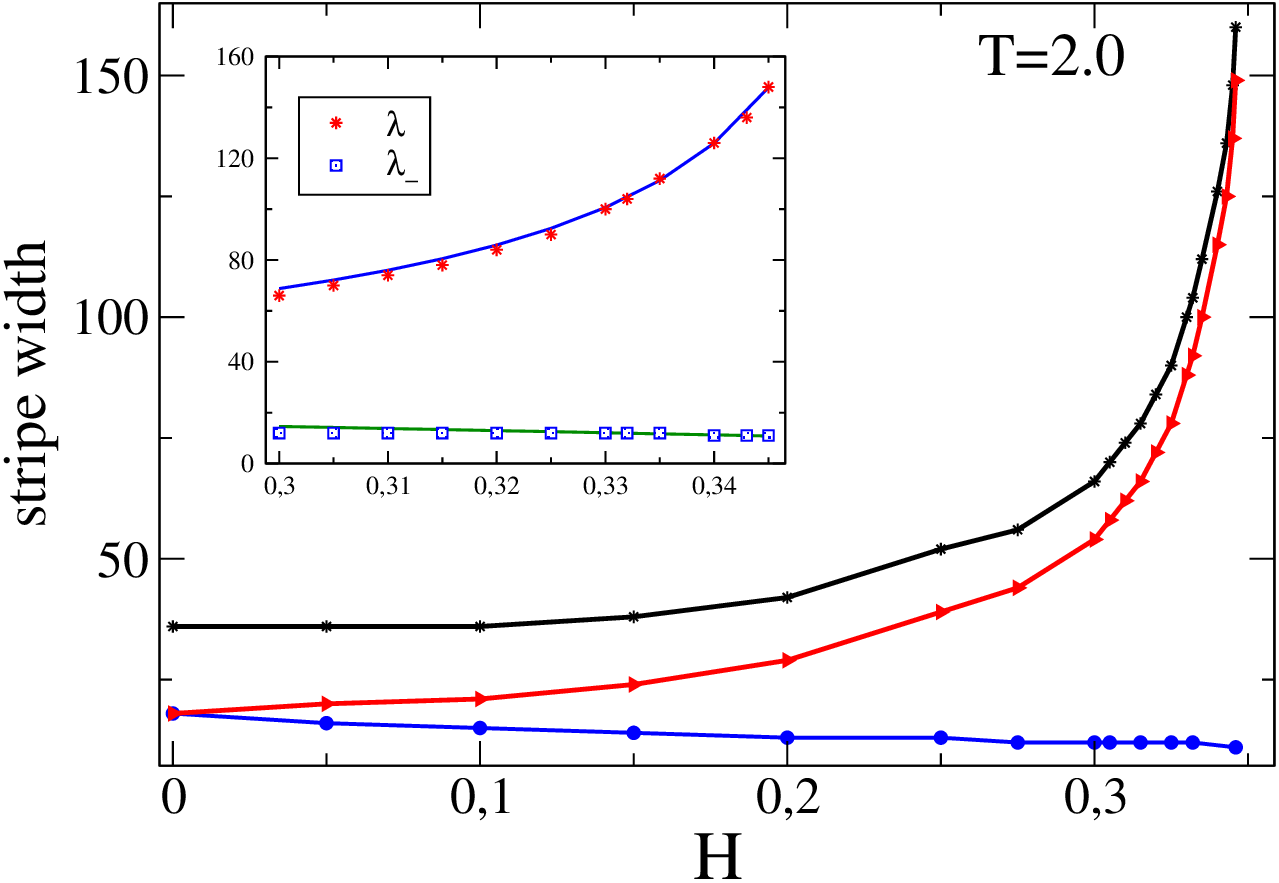}
\caption{(Color online) Stripe width $\lambda$ (black asterisk), positive component of magnetization $\lambda^+$ 
(red triangles) and asymmetry $\lambda^-$ (blue circles) as function of magnetic field for $T=2$. 
Inset: fits according to analytic predictions from reference \onlinecite{JoPaGa2013}}
\label{stripewidthT=2}
\end{figure}

All our results correspond to $\delta=6$. With this  value of $\delta$ we can capture the
physics of the problem and make the numerical analysis feasible.
At low temperatures domain walls are sharp and it is expected that all three models behave in
the same way. Of particular interest is the dependence of the stripe width, or modulation length,
for a fixed low temperature as a function of magnetic field. Because the stripes become asymmetric
under the influence of an external field, the dependence of the asymmetry parameter is also of 
interest. These functions are shown in Figure \ref{stripewidthT=2} which shows the solutions of
the mean field model at $T=2$. There is a critical
field value, which for $T=2$ is $H_c\sim 0.345$, at which the stripe width diverges. One can also
see that the positive component of the magnetization profile follows the growth of the stripe
width and also diverges, while the asymmetry (negative component) decreases very slightly from
its value at $H=0$. For $H > H_c$ an homogeneously magnetized solution has the minimum free energy.
Nevertheless, a true divergence of the stripe width is not accessible for computational limitations
on the size of the variational problem that is being solved, and then it is natural to ask
what is the precise behaviour of the stripe width near the critical field value.
Recently, T. H. Johansen et. al.~\cite{JoPaGa2013} obtained exact results for these
parameters in a model with sharp domain walls at zero temperature. In particular, they obtained that
the stripe width at $T=0$ diverges as a power law:
\beq
\lambda \propto (H-H_c)^{-1/2}
\eeq
We expect that results from the model studied here at finite but very low temperatures should behave
in the same way. In fact, a fit of our data for $T=2$ with a power law near the critical field yields a very
good agreement with the exact predictions for the sharp wall model of \onlinecite{JoPaGa2013} at $T=0$. 
The fits are shown in the inset of Figure \ref{stripewidthT=2}. An important conclusion is that the
divergence of the stripe width at a critical value of the field implies a continuous transition from
the modulated to the homogeneous phase, at variance with usual expectations. In the numerical solutions
it is possible to observe that stripes with finite widths continue to exist above the critical line,
and in this sense they represent metastable solutions with free energy larger than the homogenous one, with
a crossing at some field value. Nevertheless, this does not imply a first order transition since the
solution with minimum free energy below the critical line corresponds to one with continuously increasing
stripe width, which in fact diverges at a critical field value tending itself in a continuous way to the
homogeneous solution. This trend was also observed for any temperature along the critical line shown in
Figure \ref{phasediag}, implying
a continuous transition from the modulated to the homogeneous state at mean field level. A further evidence
is the continuous behavior of the magnetization at the critical field, as will be shown in the analysis
of section \ref{energies}.

\subsection{Phase diagrams}
\label{phases}

\begin{figure}
\centering
\includegraphics[scale=0.3]{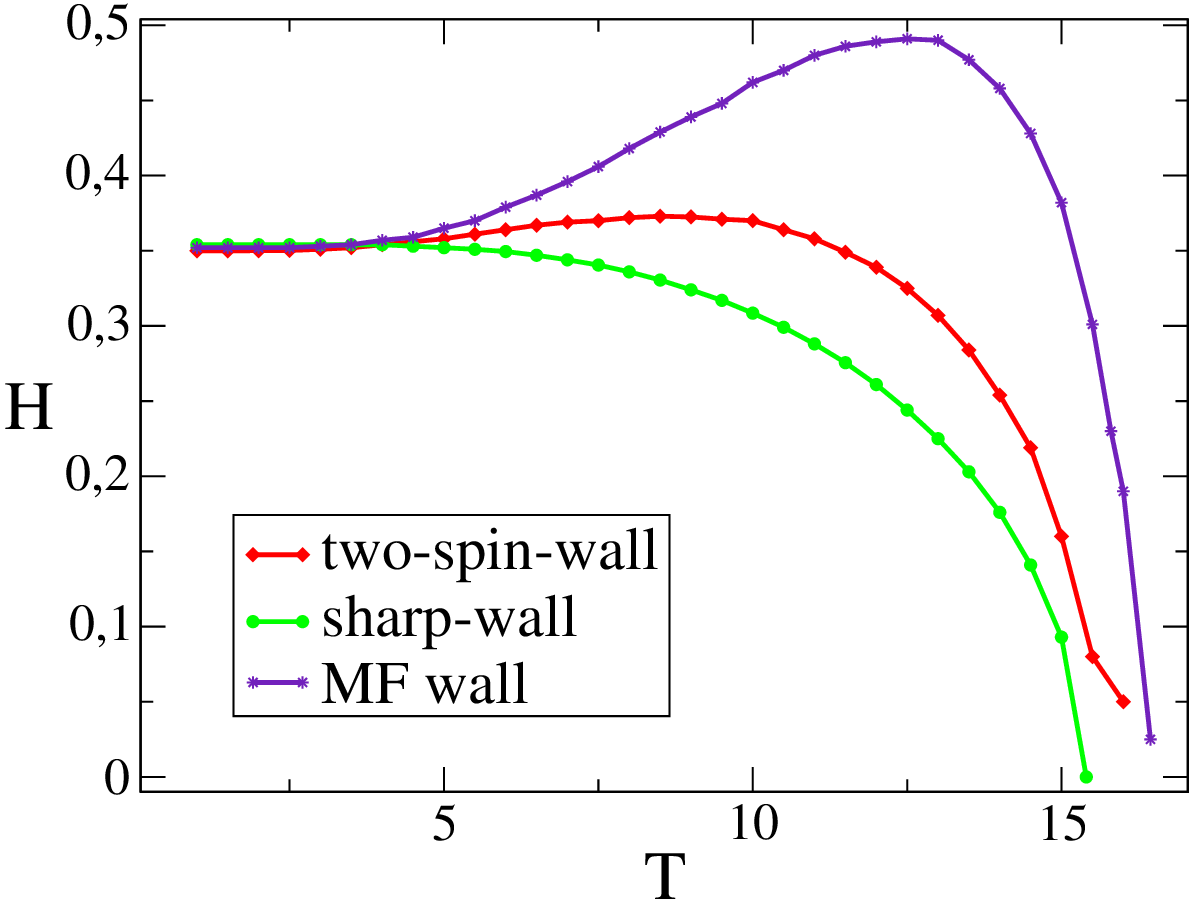}
\caption{(Color online) Field-temperature phase diagrams for the three models defined in the text.}
\label{phasediag}
\end{figure}

In order to understand the nature of the interesting reentrant behavior observed in ultrathin magnetic films with perpendicular
anisotropy, we have solved the complete mean field phase diagram of the DFIM for the asymmetric stripes solutions in the field versus
temperature plane for the three models defined above. The results are shown in Figure \ref{phasediag}. 
Below the curves for each
model the equilibrium solution is an asymmetric stripe with variable stripe width as discussed in the previous section. The
asymmetric stripe solutions compete with the homogeneous solution, which is locally stable for any $H$ at low $T$. At some
critical field $H_c$ the homogeneous solution becomes the thermodynamic one and dominates the high field section of the phase
diagram. As anticipated,
the low temperature behavior of the three models is the same. Nevertheless, at some point as $T$ grows, the critical field becomes
different for each model, signalling different behaviors. The upper curve
corresponds to the solution of the general mean field model, in which each local magnetization is considered as a variational
parameter in the minimization of the free energy. The reentrant behavior is evident. In strike contrast, the sharp-wall model
does not show signs of reentrant behavior, the critical field curve bends monotonically towards lower field values as $T$ grows,
as seen in the figure. What is the origin of the different behaviors of both models? As discussed previously, the profiles 
in both cases are very similar at low temperatures. As $T$ grows the difference must reside in the structure of the domain walls,
which is the only place where additional degrees of freedom enter the scene and affect the free energy. In fact, the sharp-wall
model has the simplest possible domain wall structure, a single discontinuity between two oppositely saturated regions. This is
the limiting case in which the domain period becomes much larger than the wall width. At mean field level, it is easy to see that
this wall will have negligible influence on the entropy of the system. At variance with this, in the full model the walls tend to
develop a non-trivial structure as temperature grows. The finite width of the walls, even at mean field level, are enough to induce
a decisive entropic contribution to the free energy of the modulated solutions. In order to further confirm if this is indeed the case,
we analyzed a model which is minimally different from the sharp-wall one, namely a model in which the walls are composed of two
sites with equal and opposite magnetization, i.e. one more degree of freedom with respect to the sharp-wall. The result is evident
in Figure \ref{phasediag}, this slight change in the structure of the wall is enough to induce a small reentrance in the phase
diagram. Our conclusion is that the structure of the domain walls is essential to the inverse symmetry breaking phenomenon seen
in ultrathin magnetic films with perpendicular anisotropy. In the last section we analyze in detail the contributions of the
energy, entropy and external fields terms to the free energy of the models.

\subsection{Free energy analysis}
\label{energies}

\begin{figure}
\centering
\includegraphics[scale=0.8]{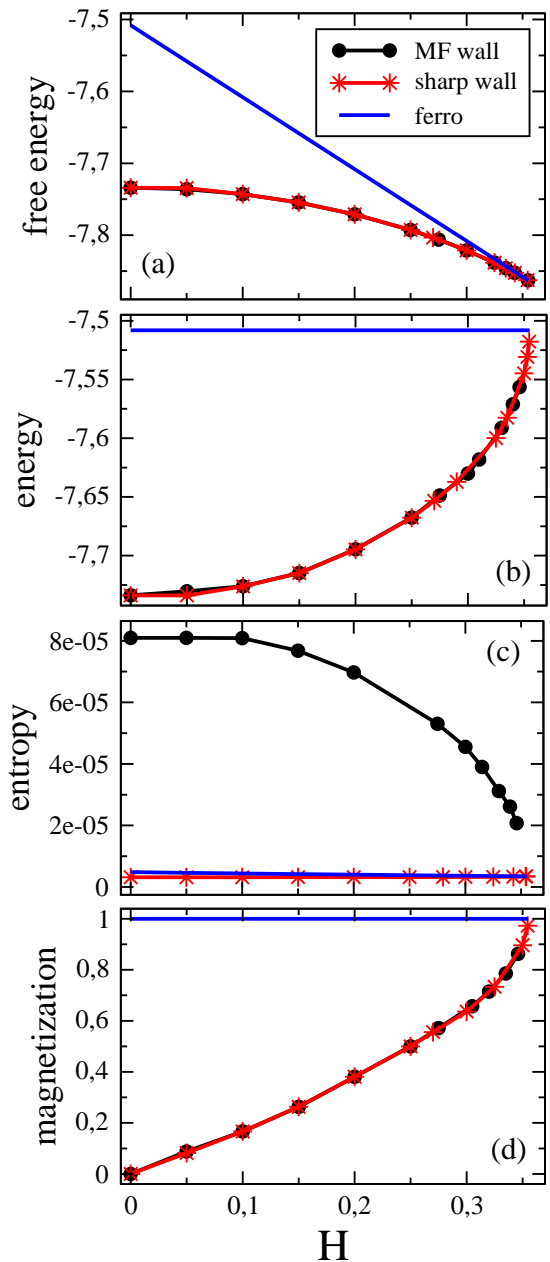}
\caption{(Color online) Free energy, energy, entropy and magnetizations as function of the applied field at $T=2$. }
\label{enerT=2} 
\end{figure}

In Figures \ref{enerT=2} and \ref{enerT=10} we show the field dependence of the free energy, energy, entropy and
magnetization for the stripe solutions of the mean field and sharp-wall models together with the homogeneous solution
at two characteristic temperatures $T=2$ and $T=10$ (see Figure \ref{phasediag}). A comparison of the free energy
curves at both temperatures confirms what was anticipated, i.e. that the full model and the sharp-wall one
behave in the same way at $T=2$, but the complete model has a lower free energy for any field value
at $T=10$, i.e. in the region under the dome where the reentrant behavior is observed. 

\subsubsection{T=2}

At $T=2$ thermal effects are negligible. This is reflected in particular in the almost zero value of the entropy.
The magnetization confirms the expectation from the exact results of the sharp-wall model of reference \onlinecite{JoPaGa2013},
the transition between the modulated and homogeneous states is continuous, signalled by the divergence of the stripe
width at a critical value of the field. Accordingly, the magnetization grows continuously from $m=0$ at $H=0$ to $m=1$
at $H=H_c$. 

\subsubsection{T=10}

\begin{figure}
\centering
\includegraphics[scale=0.8]{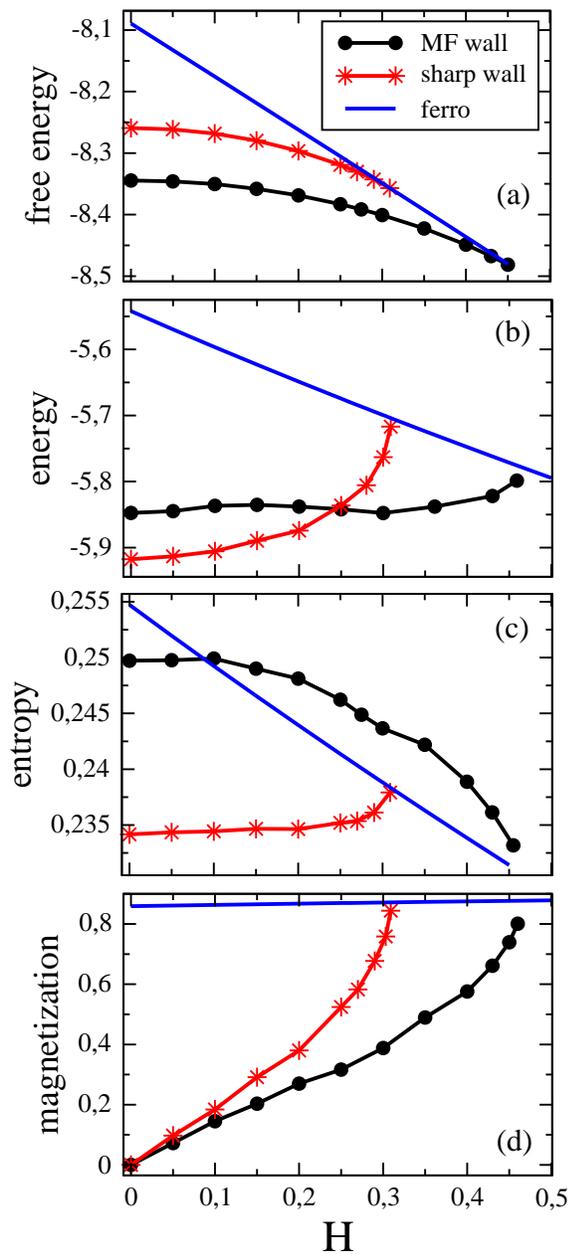}
\caption{(Color online) Free energy, energy, entropy and magnetizations as function of the applied field at $T=10$. }
\label{enerT=10} 
\end{figure}

At $T=10$ thermal effects are evident, implying a departure of both models from the common behavior seen at $T=2$.
The main difference is observed in the entropy plots of Figure \ref{enerT=10}. The mean field profiles have a notably 
higher entropy than those of the sharp-wall model. This is probably due to the extended nature of domain walls in the 
mean field model, which activates degrees of freedom not present in the sharp-wall model. Particularly for small fields, the entropic
advantage of the mean field model ensures a much lower free energy. At fields near the limit of stability of the sharp-wall
model, this has a higher magnetization than the mean field one, but this is not enough to change the balance in the free
energy, still dominated by energy and entropy contributions. As can be seen in the free energy and magnetization curves,
the transition to the homogeneous state is continuous, as for $T=2$. At $H=H_c \approx 0.46$ the stripe width diverges
similarly to what happens at low temperatures. 

Summarizing, the lower free energy of the extended domain wall model as compared to the sharp domain wall one is due to 
the combined effect of both the energy and the entropy. At low fields before the crossing of the energies of the extended domain walls
and the sharp domain wall  (see fig \ref{enerT=10} (b)), the excess of entropy manifests in a reduction of the free energy and 
after the crossing  the reduction is due to the energy contribution. As a result of these contributions the transition line 
moves to higher  fields.

In a work by Vindigni et al. \cite{ViSaPoPePo2008} the authors showed that the mean field approximation of the DFIM is 
useful for the description of the temperature behavior of the stripe width of  ultrathin Fe films epitaxially grown on Cu, 
and also that this model is adequate for the description of the domain wall profiles. 
Our results extend the application of the model to give a microscopic explanation of the reentrance behavior observed in these
systems \cite{SaRaViPe2010}.  

\newpage

\section{Conclusions}
\label{conc}

Reentrance is an interesting property related to the stability of magnetic phases.
We have shown that the inverse symmetry breaking transition in ultrathin ferromagnetic films with perpendicular anisotropy 
occurs as a result of the addtional degrees of freedom in the structure of the domain walls. We compared the phase
diagrams of three models, a mean field one with extended walls, a sharp wall model and an intermediate two-spin wall model.
When compared with a sharp domain wall scenario the extended domain has a higher
entropy reflecting the importance of the internal degrees of freedom in the reduction of the free energy. Furthermore, 
our results show that with sharp domain walls the inverse transition is absent. On the other hand, the structure of the domain 
walls does not affect the dependence of the period of the stripes as function of the applied field. We realized that for temperatures 
before the bump in the phase diagram, the stripe period is well described according to a zero temperature model which assumes sharp domain
walls. This implies that in this range of temperatures the transition between the ferromagnetic and the stripe phase 
is continuous in the mean field approximation. Furthermore, our results for the mean field model indicate that the whole critical line is a
line of continuous transitions. 
The sensitivity of the inverse transition to the internal degrees of freedom of the domain walls was tested through the addition of one
degree of freedom to the sharp wall model. The results show that this is enough to induce an inverse transition, completely absent in the
sharp wall model. The mechanism of ISB can be summarized as
follow: in the low temperature range the domain wall profiles are sharp and the critical field for reaching the ferromagnetic phase
is nearly constant. As the temperature increases the domain walls aquire some finite width and structure and then the entropy increases
inducing a lowering of the free energy and hence a higher field is needed to enter into the homogeneous ferromagnetic phase. 

It would be interesting to test this conclusions in models with realistic domain walls and quantify experimentally the extension of the reentrant
phenomenon in thin films and its influence on the stability of magnetic domains.

\acknowledgments

This work was partially supported by CAPES/SPU through grant PPCP007/2011, CONICET, SeCyT and Universidad Nacional de C\'ordoba
(Argentina) and CNPq (Brazil).
%
%

\end{document}